\documentclass{WileyMSP-template}

\usepackage[superscript]{cite}

\usepackage[T1]{fontenc}
\usepackage{siunitx}  
\usepackage{amsmath}
\usepackage{graphicx}
\usepackage{xcolor}
\usepackage{upgreek}
\usepackage{wrapfig}

\begin{document}

\pagestyle{fancy}
\rhead{\includegraphics[width=2.5cm]{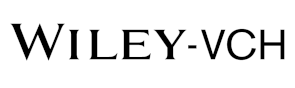}}

\title{Interlayer-catalyst method for metal assisted chemical etching of ultra-high aspect ratio silicon nanostructures}

\maketitle

\author{Bryan Peter Jost Benz*,}
\author{Marco Stampanoni}
\author{and Lucia Romano.}

\begin{affiliations}
B.P.J. Benz\\
PSI Center for Photon Science, Paul Scherrer Institut, 5232 Villigen PSI, Switzerland\\
Department of Physics and Swiss Nanoscience Institute, University of Basel, 4056 Basel, Switzerland\\
Email address: bryan.benz@psi.ch

Prof. Dr. M. Stampanoni, Dr. L. Romano\\
PSI Center for Photon Science, Paul Scherrer Institut, 5232 Villigen PSI, Switzerland\\
Institute for Biomedical Engineering, University and ETH Zürich, 8092 Zürich, Switzerland\\

\end{affiliations}


\begin{abstract}

Reliable and precise etching of silicon nanostructures with ultra-high aspect ratios is required in many fields. Metal assisted chemical etching (MacEtch) in HF vapor and O$_{2}$ is a plasma-free etching method that attracts considerable attention owing to its ability to create smooth, high aspect ratio nanostructures. MacEtch understanding and applications are limited by low fidelity and inconsistent pattern transfer from the catalyst layer to the silicon substrate. The locally constrained electrochemical interactions at the catalyst site make MacEtch particularly sensitive to catalyst contamination, which reduces the reaction rate and pins the catalyst during etching. Removing contaminants is essential to improve pattern transfer for reliable processes over a larger area and with a higher aspect ratio. Physically separating the main source of carbon, the resist, from the catalyst with a sacrificial and functional interlayer solves this issue. The interlayer separates the resist and the catalyst and allows thorough cleaning of the substrate before catalyst deposition. The resulting clean catalyst enables the fabrication of dense (50\% patterned area) high aspect ratio (>250:1) nanostructures. Different interlayer materials (Cr, Al$_{2}$O$_{3}$ and SiO$_{2}$) and two patterning approaches are presented, showcasing etching of various high aspect ratio nanostructures, such as X-ray optics.  

\end{abstract}

\section{Introduction}
The micro- and nanoscale domains benefit significantly from structures with high aspect ratio (HAR) and precise feature control in semiconductor materials, especially in silicon. Consequently, etching methods\cite{RN2650} played a crucial technological role in the device miniaturization roadmap and have attracted considerable attention in the nanofabrication era. Despite their ability to achieve very high aspect ratios of 100:1, wet-etch processes based on alkaline solutions \cite{RN1840} and electrochemical etching have distinct limitations.\cite{RN3159} Etching in alkaline solutions is geometrically constrained by anisotropy based on crystalline direction,\cite{RN314} while electrochemical etching is highly dependent on substrate doping and less flexible for arbitrary pattern transfer than deep reactive ion etching despite being competitive in speed with etch rates of over 8 $\upmu$m min$^{-1}$ .\cite{RN3159, RN621} In plasma etching processes, the etching anisotropy relies on the formation of protective passivation layers on the feature sidewalls. The Bosch process consists of a cyclic repetition of an etching step using SF$_{6}$ plasma, followed by a fluorocarbon plasma step to passivate the trench sidewalls, and works well for silicon aspect ratio of 20:1 and feature size in the range of 1-100 $\upmu$m. Cryogenic (-100 °C) reactive ion etching can achieve aspect ratios of approximately 120:1 for sub-micrometer trench widths.\cite{RN445} O$_{2}$  passivation instead of fluorocarbon at room temperature allows nanostructuring capability but extremely slow etching rate.\cite{RN2065} Metal assisted chemical etching (MacEtch)\cite{RN312,RN15,RN29} emerged as an alternative technique to conventional methods, with unique advantages of plasma-free and anisotropic nature. MacEtch can etch materials with high anisotropy, smooth sidewalls, and a large range of  feature sizes, from 1 nm to 100s of micrometers.\cite{RN221,RN78} The associated research encompasses a wide range of applications, including anti-reflective surfaces,\cite{RN15} photovoltaics,\cite{RN62} energy harvesting and storage,\cite{RN257} sensors,\cite{RN60} bio-interfaces,\cite{RN1594} fingerprints \cite{https://doi.org/10.1002/smll.202500878}, vias\cite{7159048} and X-ray optics.\cite{RN295} The substrates available for etching via MacEtch are diverse, and include Ge, GaAs, GaN, InP, and predominantly Si.\cite{RN312} The MacEtch process involves a local electrochemical reaction occurring at the catalyst site. During etching of a silicon substrate, the catalyst sinks into the substrate. The anisotropic nature of etching originates because it occurs predominantly at the interface of silicon in contact with the catalyst layer.\cite{RN2535}  For precise HAR pattern transfer, the catalyst should preserve its shape and activity during the whole etching process. Etch speed differences might occur as a consequence of local pinning of the catalyst\cite{RN150} tearing the catalyst apart and leading to uncontrolled etching. Such local pinning manifests as relevant distortions of the etching direction, which is usually intended to be vertical to the substrate. Under these conditions, MacEtch does not retain the original pattern and instead produces defects and random microporosity in the Si substrate. This poses a critical challenge for the consistent manufacturing of nanostructures with vertical sidewalls and high fidelity pattern transfer. Pattern integrity remains one of the main challenges for MacEtch, preventing its widespread application in various fields.\cite{RN173,znati_wharwood_tezanos_li_mohseni_2024}  The patterning approach presented in this work avoids the catalyst contamination, which is at the basis of the pattern degradation during MacEtch, in the sub-micrometer regime and in the presence of electron beam lithography (EBL). The improved patterning demonstrates the relevance of controlled catalyst preparation and reveals the new capabilities of MacEtch in the gas phase for semiconductor manufacturing and X-ray optics. 

\subsection{MacEtch in gas phase }

MacEtch with reactants and by-products in the gas phase\cite{RN295,RN57} eliminates the drying step after etching in liquid solutions.\cite{RN255} In essence, MacEtch in the gas phase is analogous to its liquid counterpart:\cite{RN236} an oxidant, either O$_{2}$ as presented here or O$_{3}$,\cite{RN2174} is reduced at the catalyst side by extracting electrons from Si or equivalently injecting positive charge carriers in the underlying substrate, causing the oxidation of silicon;\cite{RN208} the etchant, hydrofluoric-acid (HF), attacks oxidized silicon, leading to Si removal at the interface with the catalyst, with the consequent sinking of the catalyst into the substrate. It has been shown that the microdynamic behavior of etching depends on the chemical nature of the catalyst.\cite{RN2667} In gas, the oxidant of choice is O$_{2}$ and the catalyst is Pt. Other options are available, but this combination has achieved ultra high aspect ratio structures.\cite{RN295}
The oxygen reduction on Pt is regulated by Equation (1)\cite{XIANG2021107039} and mainly occurs at the top surface of the catalyst, whereas direct Si etching follows Equation (2) and mainly occurs at the catalyst/substrate interface. The balanced Equation (3) is obtained by combining equations (1) and (2):
\begin{align}
O_{2(g)} + 4H^{+}  + 4e^{-}_{(Si)} &\rightarrow 2H_{2}O_{(l)}\\
Si_{(s)}+4h^+_{(Si)}+4HF_{(g)} &\rightarrow SiF_{4(g)} + 4H^+      \\
Si_{(s)}+O_{2(g)}+4HF_{(g)} &\rightarrow 2H_{2}O_{(l)}+SiF_{4(g)} 
\end{align}
where (s), (l), and (g) indicate the solid, liquid, and gas states of the molecules, respectively.  The process is usually realized at temperatures higher than 45 °C to reduce the condensation of liquid (evaporated HF from the liquid or water produced as a by-product) on the surface of etched silicon nanostructures. If the rate of charge carrier injection from the oxidant reduction exceeds the rate of atom removal by the etchant, the excess holes diffuse away from the catalyst site, causing undesirable etching at locations not in direct contact with the catalyst. For a successful MacEtch process, hole injection and mass transfer of the reactants must occur uniformly at every catalyst site and during the full process time. If this is not the case, etch speed differences may cause pattern distortions and degrade pattern transfer fidelity. Because reaction (1) occurs at the catalyst surface and reaction (2) occurs at the interface between silicon and the metal surface, the catalyst surface and interface quality are crucial.\cite{RN323} Impurities clogging the catalyst surface or the interface between the catalyst and the Si substrate might limit the charge carrier production and injection. Pinning of the catalyst has been reported in MacEtch to induce 3D spiralizing movement of the catalyst, \cite{RN150,RN151} whereas a cycled variation in oxidant concentration can induce zigzag structures.\cite{RN25} Controlling the etching direction is the key to achieve the highest fidelity of pattern transfer in HAR silicon nanostructures, with the best catalyst path movement being perpendicular to the substrate. Previous work demonstrated that in the gas phase, the reactants can be supplied continuously from an - considering the timescale - infinite reservoir: HF evaporates from a liquid not in direct contact with the catalyst-patterned sample, and O$_{2}$ is supplied by air continuously entering the etching chamber. In the gas phase, the oxidant concentration can be much smaller than that of a typical wet-MacEtch, so that the porosity of the etched structures, caused by excess holes, can be drastically reduced.\cite{RN44} A local variation in hole injection can cause a local variation in the etching speed and direction with a consequent distortion in the vertical path of the catalyst movement. Owing to the lower concentration of reactants \cite{RN44} in the vapor with respect to the liquid solutions, any local imperfection - or the presence of impurities that can alter the MacEtch reaction - is more critical in the gas phase. 

\subsection{Patterning issues for nanostructures}

With the progressive movement of the catalyst pattern down into the substrate, the diffusion of reactants through the etched structure becomes the limiting factor for supplying a constant and uniform amount of reactant at the catalyst side. We demonstrated that nanowires with aspect ratios of up to 10'000:1\cite{RN295} can be realized by self-assembled catalyst patterns, indicating that reactant diffusion is not the limiting factor for this aspect ratio. A much lower aspect ratio has been reported with lithographically patterned nanowires and the same etching method\cite{SHI2023107311} indicating that pattern transfer is crucial and responsible for introducing defects, preventing the realization of HAR nanostructures with a lithographic design. A typical method for lithographically patterning a catalyst is through a lift-off process. In this process, a catalyst is deposited on a developed resist, then the resist is removed by solvents lifting off the catalyst so that only the catalyst deposited on the resist-free area is left on the silicon substrate.  Oxygen plasma treatment prior to deposition is required to clean the resist residues at the cost of degrading the profile of the patterned resist.\cite{RN363} Noble metals are usually deposited by physical evaporation, implying a substantial warming of the resist surface during the deposition, which might result in resist reflow or decomposition, leading to a poor yield of the lift-off process (see Supporting Information),\cite{RN2627} especially for feature sizes in the submicron regime. Pt also exhibits poor adhesion to Si, which often leads to delamination, especially in solvents.\cite{vervuurt_sharma_jiao_kessels_bol_2016,al-shareef_dimos_tuttle_raymond_1997} Standard adhesion layers, such as Cr or Ti, might alter the interface contact of catalyst and substrate, resulting incompatible to MacEtch, especially in the gas-phase. Several methods have been developed to overcome the poor lift-off yield, including direct electrochemical nanoimprinting,\cite{RN14,RN323} flow-enabled self assembly of block co-polymers, \cite{RN225} resist masking,\cite{RN326} nanoimprint with an undercut layer,\cite{RN40, RN158} patterning of a seed layer and subsequent electroplating of the catalyst,\cite{RN2627} shadow masking the deposition of the catalyst\cite{RN35} and stamping.\cite{RN202} While these methods manage to circumvent lift-off, they either do not allow for patterning freedom at the nanoscale (electrochemical nano-imprint, flow-enabled self assembly, masking) or fail to effectively separate the resist from the catalyst (resist masking , nanoimprint, electroplating, stamping). EBL remains the most widely used method for patterning catalysts for nanostructures. We observed that MacEtch of nanostructures created with a standard lift-off approach often leads to inconsistent and random etching, presenting local distortions and tilting of the catalyst pattern at the beginning of the process, making it extremely rare to realize HAR with good quality. In contrast, structures created without EBL do not have these issues, showing very robust patterning integrity even after a long etching time with an extremely high reproducibility yield over a very large range of etching conditions.\cite{RN295} These observations suggest that EBL is the main cause of the poor pattern transfer fidelity in MacEtch. The culprit is hypothesized to be the contamination of the surface that remains after the development of EBL resist and before catalyst deposition. The presence of resist residues is well known in EBL \cite{Mao_2025} and is critical for any etching process. Electron-beam-deposited carbon has been demonstrated to act as a mask for MacEtch,\cite{RN77}  highlighting the criticality of EBL for MacEtch with respect to other techniques, such as UV lithography and nanoimprinting. XPS and SEM images support the hypothesis of a larger carbon contamination of the surface after EBL, even after oxygen plasma cleaning (see Supporting Information). Because surface cleaning after EBL patterning is limited to preserve the resist profile, we solved the issue by physically separating the resist from the catalyst layer during EBL patterning.
In this study, we introduced an additional layer, that is inert for the MacEtch process, henceforth called "interlayer", between the resist layer and the catalyst layer. The interlayer serves as a medium to pattern the catalyst on clean Si by allowing robust substrate cleaning before catalyst deposition.  \textbf{Figure \ref{fig:interlayer_etching}} and \textbf{\ref{fig:interlayer_LO}} show the essential workflows of the proposed approaches. Examples of dense nanostructures, such as optical elements for X-ray imaging applications\cite{RN709} with feature sizes in the range of 100-400 nm and aspect ratios higher than 250:1, are realized by MacEtch in the gas phase. 

\newpage
\section{Results and discussion}

The presence of an interlayer requires an additional step to transfer the pattern from the EBL resist to the catalyst layer. We present two distinct methods to prepare the pattern: etching (I-Et) and lift-off (I-Lo) of the interlayer material. The platinum thickness was always 12 nm, whereas the interlayer thickness was varied between 20 and 80 nm. The structures are etched using MacEtch in gas phase, we described the etching optimization procedure on selected examples in the last section. The pattern design and technical details of the processing are described in the Materials and Methods section. Standard lift-off and plasma etching approaches to prepare the catalyst pattern are reported for comparison in the Supporting Information.

\subsection{Interlayer pattern transfer by etching}
\begin{figure}[h]
    \centering
    \includegraphics[width=1\textwidth]{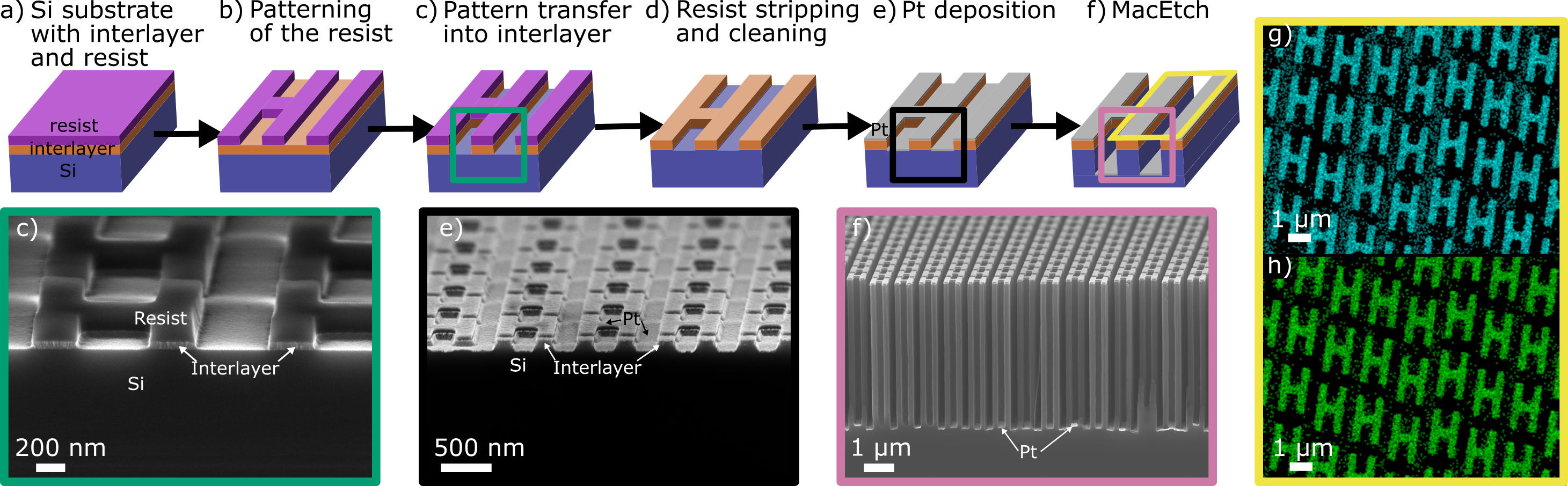}
    \caption{a)-f) show the schematic process flow of the I-Et approach to patterning. The text describes the process flow. Cross-sectional SEM images are provided for three selected steps: (c) after pattern transfer (the Cr interlayer 50 nm-thick is clearly distinguished from the patterned EBL resist); (e) after Pt deposition (the 12 nm Pt layer is clearly distinguished from the Cr interlayer underneath); and (f) after MacEtch (Pt is at the bottom of the etched silicon structures). Top-down EDX images are provided for Pt cataylst signal (g) and the Cr interlayer signal (h) after MacEtch in the same area, showing Cr/Pt top coating of the etched Si structures.}
    \label{fig:interlayer_etching}  
\end{figure}

Figure \ref{fig:interlayer_etching} illustrates the interlayer etching (I-Et) process flow and its characterization in cross-section scanning electron microscopy (SEM) and Energy Dispersive X-ray analysis (EDX). First, a Si substrate is coated with an interlayer material, here Cr, and a resist. The resist is patterned via EBL, and the resist pattern is then transferred to the interlayer via plasma etching (Figure \ref{fig:interlayer_etching}.c).

This method facilitates the complete removal of the resist layer prior to the deposition of the catalyst. Robust cleaning can be performed after pattern transfer to the interlayer using oxygen plasma and chemical agents. Cr is an attractive choice for the interlayer because it has been shown to act as a MacEtch block \cite{RN11, RN954, RN638, RN1633,RN40} and is chemically stable during the process used to develop and strip the resist. This enabled the complete removal of the resist without significantly damaging the interlayer. Cr exhibits good stability in HF vapor. A cross-sectional SEM image (Figure \ref{fig:interlayer_etching}.f) shows the results of the I-Et method with 50 nm of Cr after 15 min of MacEtch at 55 °C above HF [25\%]. The resulting Si structure is an H-bar pillar with a linewidth of 200 nm and a height of 4.91 $\pm$ 0.04 $\upmu$m. A clean interface between the catalyst and substrate allows for consistent and straight etching (Figure \ref{fig:interlayer_etching}.f). 
EDX data (Figure \ref{fig:interlayer_etching}.g,h) comes from a 400 nm linewidth structure (I-Et, 50 nm Cr) after 15 min of MacEtch at 50 °C above HF [25\%]. The etched depth of the structure was measured by SEM and is 7.5 $\pm$ 0.18 $\upmu$m. EDX maps indicate that Cr and Pt are co-present on the tips of the Si etched structures after MacEtch. The SEM cross-sectional image (Figure \ref{fig:interlayer_etching}.k) shows the efficacy of the Cr layer in obstructing the MacEtch process, as evidenced by the negligible etching observed beneath the Cr interlayer. This finding demonstrates that Cr has MacEtch-blocking capability not only in wet-MacEtch with H$_2$O$_2$ as the oxidant\cite{RN11} but also in the gas phase with O$_2$.

\subsection{Interlayer pattern transfer by lift-off}
\begin{figure}[h]
    \centering
    \includegraphics[width=1\textwidth]{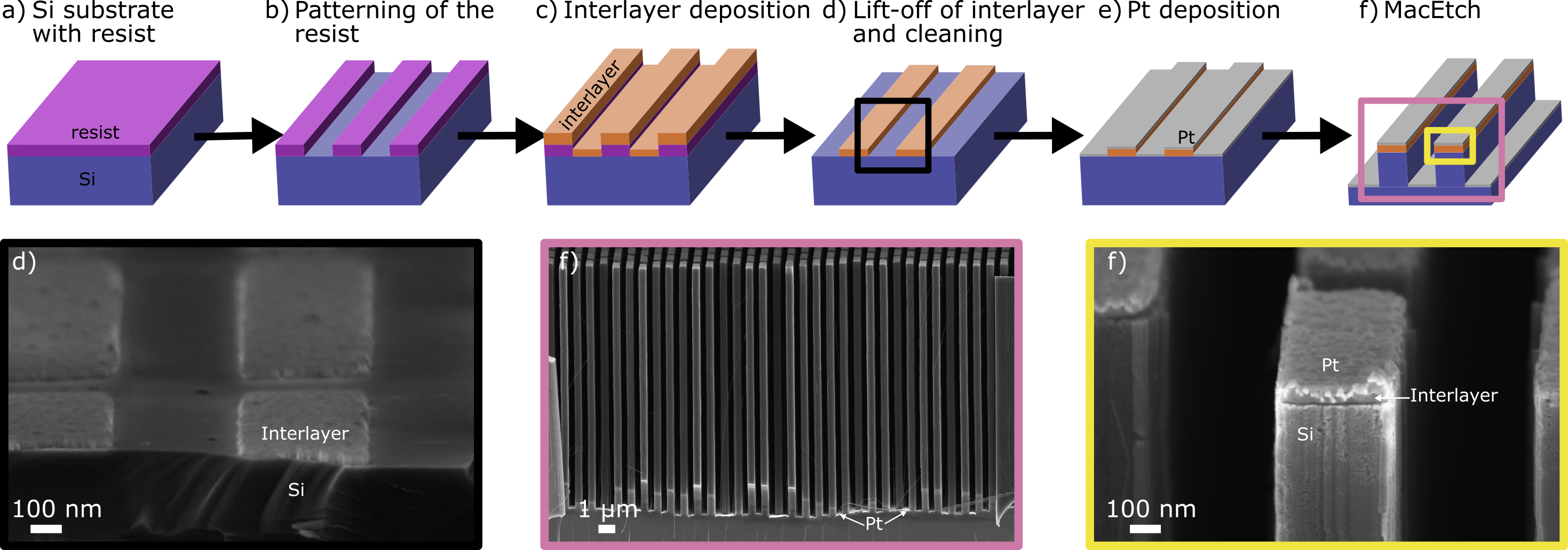}
    \caption{Schematic process flow of the I-Lo approach to patterning. Cross-sectional SEM images of selected steps with 50 nm Al$_2$O$_3$ as the interlayer: d) after lift-off of interlayer (the interlayer pattern is visible in SEM contrast with respect to the Si substrate); f) after MacEtch in low resolution (Pt catalyst at the bottom of the etched structures) and high resolution (the interlayer and Pt are distinguishable on the top of the etched Si lamella). }
    \label{fig:interlayer_LO}  
\end{figure}

The lifting-off of the interlayer provides an alternative for patterning it. This method avoids the use of plasma processes by employing only liquid chemicals. The corresponding interlayer lift-off (I-Lo) process flow is illustrated in Figure \ref{fig:interlayer_LO}. First, the pattern is written on a resist layer coating a Si substrate; the interlayer material is then deposited by physical evaporation; the interlayer is subsequently removed by lift-off using a chemical that dissolves the resist layer; after a robust cleaning, the catalyst is deposited; and finally, the pattern is transferred into the Si substrate by MacEtch.

Analogous to the I-Et approach, the I-Lo approach facilitates the removal of the resist before catalyst deposition. The process flow enables comprehensive and robust cleaning using oxygen plasma prior to Pt deposition, without the side effect of pattern deterioration, such as in the case of EBL resist. Using the I-Lo method, materials that are difficult to etch by the plasma process can be used as interlayers. 
As an option for an additional interlayer material, we present here Al$_2$O$_3$. Similar to Cr, Al$_2$O$_3$ is resistant to oxygen plasma. As an oxide layer, it is compatible with a larger selection of processes and applications and is CMOS-compatible. Previous studies have used Al$_2$O$_3$ as a stopping layer for HF vapor etching of SiO$_2$.\cite{RN368} In our study, we see that it is also able to block MacEtch. In liquid MacEtch, aqueous HF rapidly consumes Al$_2$O$_3$.\cite{williams_gupta_wasilik_2003,RN369} In contrast, in gas-MacEtch, Al$_2$O$_3$ will survive, as HF vapor does not attack Al$_2$O$_3$ in a significant manner.\cite{williams_gupta_wasilik_2003,bakke2005etch} Figure \ref{fig:interlayer_LO} reports the process flow with Al$_2$O$_3$ interlayer, nanostructures with a linewidth of 400 nm were etched for 30 min at 60 °C above HF [50\%] to reach an etching depth of 14.93 $\pm$ 0.17 $\upmu$m (Figure \ref{fig:interlayer_LO}.f). The metal oxide tip (Pt-Al$_2$O$_3$-Si) is visible in the high-magnification SEM image of the top part of the etched grating (Figure \ref{fig:interlayer_LO}.f), no significant undercut or erosion is observed on the top surface, as expected for anisotropic etching by MacEtch in the gas phase, where the injection of charge carriers by oxidant reduction can be tuned much better than in the liquid reaction. With respect to Cr, Al$_2$O$_3$ is more compatible to thermal treatments, such as the one required for Pt dewetting to create a porous catalyst layer.\cite{RN295} 

It is worth nothing that in I-Lo the unpatterned area will produce active Pt area, while in I-Et the unpatterned area has blocked and therefore inactive Pt. This makes a direct comparison of the two methods with the same etching conditions impractical. Thus, the I-Lo approach represents a quasi-inverse lithographic process compared to the I-Et approach. I-Lo and I-Et can be combined as a function of the required pattern, providing complementary designs for a more efficient lithography process. In direct-writing lithography, the proper choice of patterning can reduce the processing time, particularly for large-area patterns. The same EBL resist can be used for both methods, which simplifies the writing and enables the use of the preferred resist for both approaches, irrespective of the positive or negative tone of the resist.
Compared with the I-Et method, I-Lo provides a universal solution that does not require selective etching of multiple layers. This enables the use of any stack of materials that survives exposure to HF vapor and has one layer of blocking MacEtch. In principle, any material with a different MacEtch behavior than that of the material used as the main catalyst is a possible candidate for the interlayer, including those that catalyze the MacEtch process, provided that the interlayer material has a different etching rate with respect to the main catalyst. This method enables the use of a stack comprising one material that acts as a MacEtch blocker and another that imparts functionality to the stack. Top coating materials can be deposited prior to etching as part of the lithographic process, facilitating the streamlining of the process because no second patterning step before etching or deposition after etching is required.

\subsection{Enhanced interlayer profile}
\begin{figure}
\centering
    \includegraphics[width=0.45\textwidth]{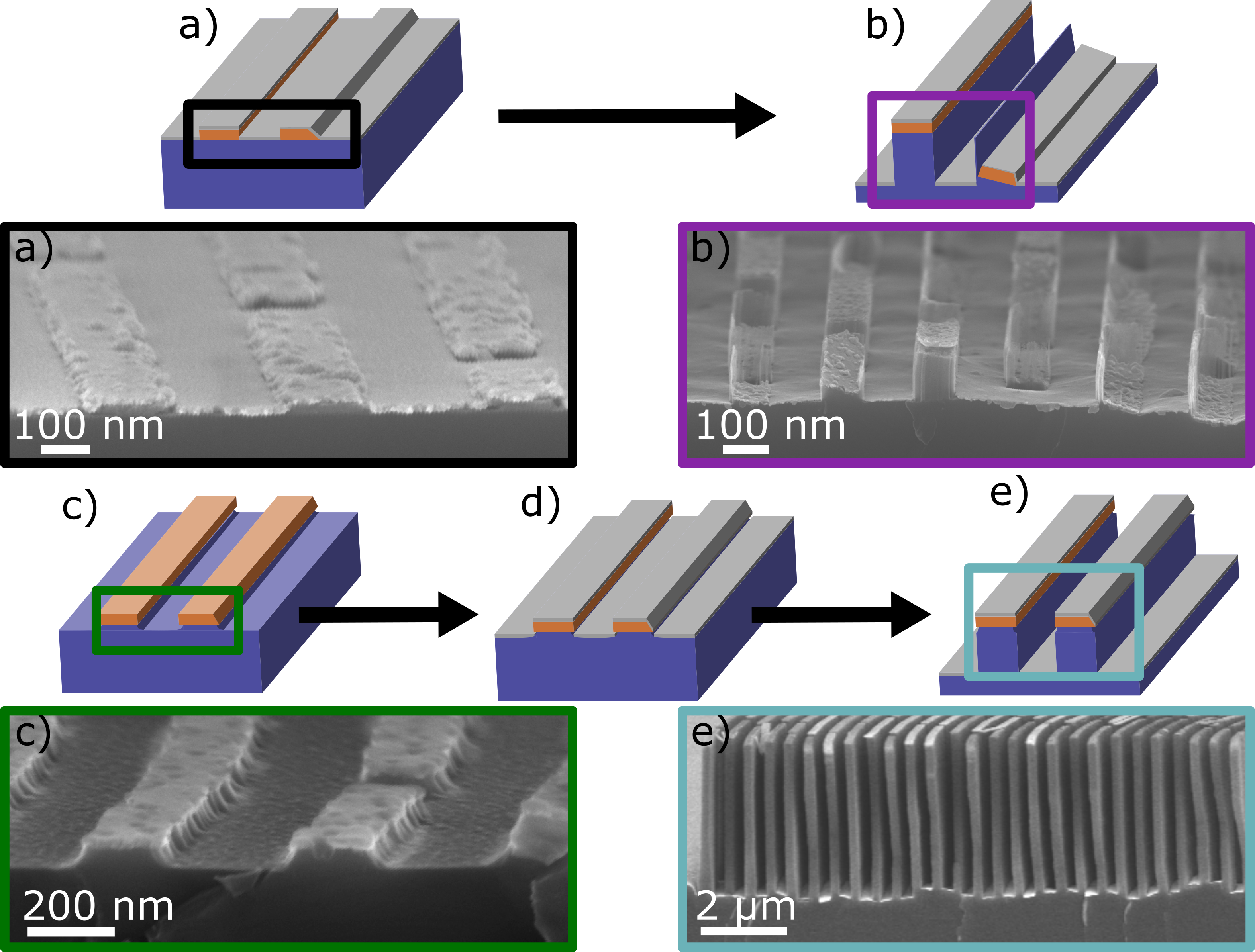}
    \caption{ A schematic showing the issue in pattern transfer arising from a poor interlayer profile (a-b), solved by 20 nm plasma etching of Si using the interlayer as hard mask (c-e). The supporting cross-sectional SEM images show select steps in the process.}
    \label{fig:profile}  
\end{figure}

The success of the interlayer method relies on effective catalyst separation during deposition. \textbf{Figure \ref{fig:profile}}  shows an example of poor pattern transfer with defects, high porosity, and/or destroyed tops, as caused by incomplete catalyst/interlayer separation. During MacEtch, the catalyst is either blocked or sinks and drags the top with it. In both cases, charge carriers are injected from the side to etch the structure laterally. If the catalyst sinks with the top, one side of the structure remains, leaving behind characteristic fences and a sunken top (Figure \ref{fig:profile}.b). To avoid these defects, the interlayer should have straight sidewalls or an undercut before catalyst deposition.\cite{RN40} Figure \ref{fig:profile}.c presents the effect of an additional isotropic etching of Si to create an undercut below the patterned interlayer before the Pt deposition. Using SF$_6$/O$_2$ plasma treatment, we etched 20 nm of Si using the interlayer as a hard-mask. Such a small artificial undercut leaves the top of the structure with minimal damage and supports catalyst disconnection. The example structure with a 30 nm Cr interlayer was etched for 10 min at 55 °C above HF [50\%] (Figure \ref{fig:profile}.e) and presents a more uniform and regular pattern transfer in the substrate with respect to the same sample without undercut (see Figure  \ref{fig:profile}.b).

\begin{figure}
    \centering
    \includegraphics[width=\textwidth]{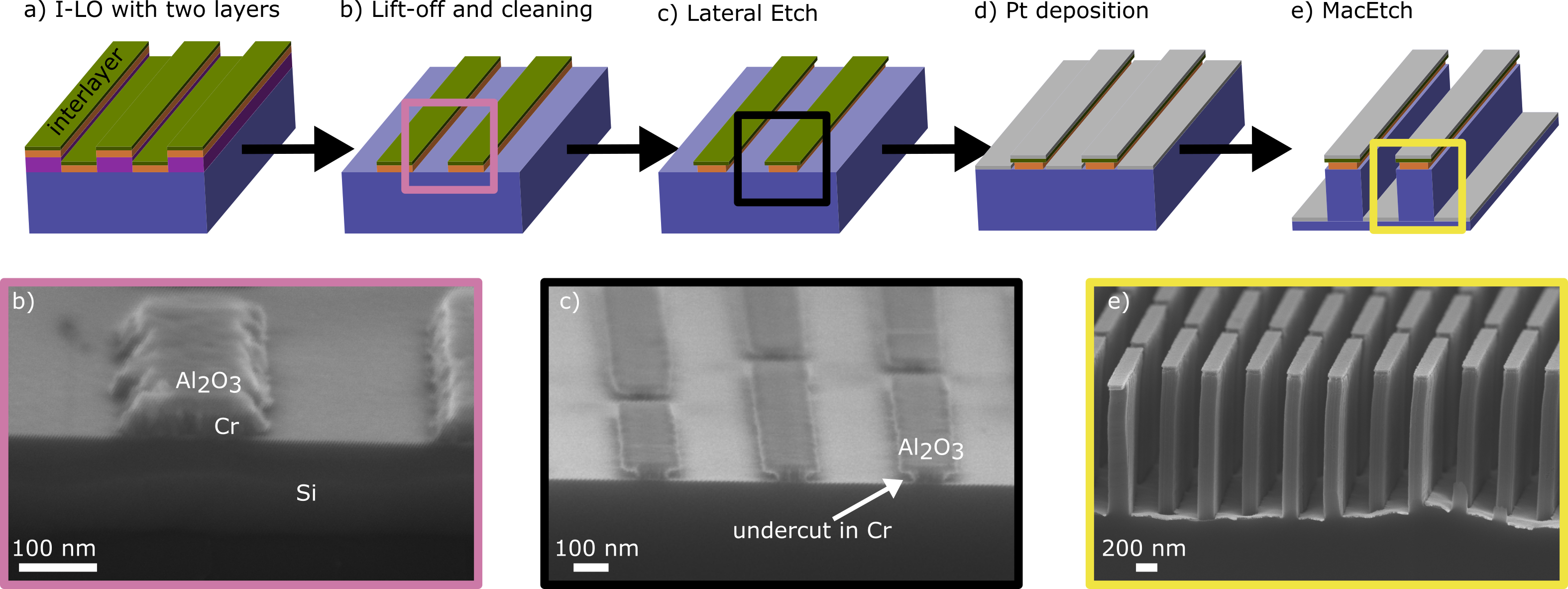}
    \caption{ A schematic showing an advanced process using a bilayer as the interlayer to create an undercut. The scheme proceeds as I-Lo, followed by a selective lateral etch of the lower layer before depositing Pt.  Cross-sectional SEM images of  selected steps are provided for the process with Cr/Al$_2$O$_3$ as the interlayer: b) after lift-off and cleaning (the interlayer profile has inclined sidewalls preventing a good catalyst disconnection); c) after lateral etch to create the undercut in the Cr layer; c) after MacEtch, showing a regular profile of Pt catalyst.}
    \label{fig:bilayer}  
\end{figure}

An undercut can also be created using a multilayer consisting of two materials, with the ability to etch one selectively over the other to create an undercut layer. Here, we report an example with a double layer of Cr at the bottom and Al$_2$O$_3$ on top. The process is outlined in \textbf{Figure \ref{fig:bilayer}} for the I–Lo approach, with supporting cross-sectional SEM images. It starts with the exposure of the resist, followed by the deposition of a bilayer of Cr/Al$_2$O$_3$ (30 nm/20 nm) as the interlayer (Figure \ref{fig:bilayer}.a)  and subsequent lift-off (Figure \ref{fig:bilayer}.b). In this step, the sample is cleaned from resist residuals. The undercut is then created by immersing the sample in a diluted solution of Cr etchant (see Materials and Methods for details) to achieve lateral etch rates in the nm s$^{-1}$ regime. The result is a clearly defined undercut in Cr (Figure \ref{fig:bilayer}.c). A plasma process to undercut Cr is also available (see Supporting Information).  The catalyst is then deposited (Figure \ref{fig:bilayer}.d), and MacEtch performed with this structure (Figure \ref{fig:bilayer}.e). The cross-sectional SEM image shows the result of a grating etched at 50 °C for 15 min above HF[37.5\%]. We also realized an undercut in Al$_2$O$_3$ using SiO$_2$ as a mask (see Table 1 in the Materials and Methods section). The undercut poses an obvious limit to the smallest achievable feature size, which cannot be smaller than twice the undercut length.

\subsection{MacEtch processsing }

\begin{figure}
    \centering
    \includegraphics[width=1\textwidth]{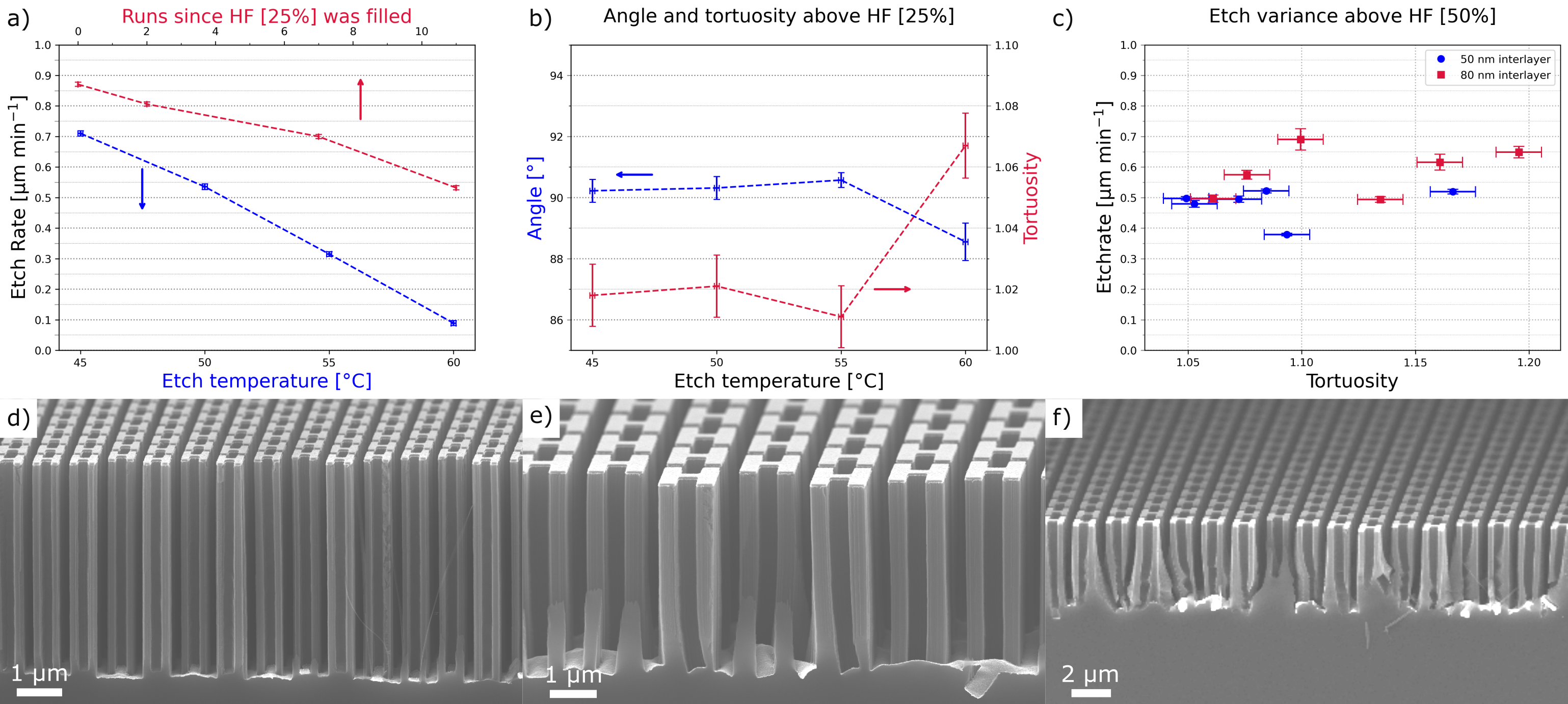}
    \caption{ H-bar structures with 200 nm linewidth (I-Et, 50 nm Cr, 15 min HF[25\%]): a) etch rate as a function of sample temperature and number of runs (opening the pot and letting the vapor out-diffuse); b) etching angle of the lines with respect to the substrate and catalyst tortuosity ($\tau$) as a function of temperature; SEM in cross-section of three examples with different quality of the catalyst at the bottom of the etched trenches: d) good quality ($\tau$=1.03), e) higher tortuosity ($\tau$=1.05) without tears, and f) catalyst with isolated motions and breaks ($\tau$=1.23). H-bar structures with 200 nm linewidth (I-Lo, 30 min HF[50\%], 55 °C) with different Al$_2$O$_3$/SiO$_2$ interlayer thickness, showing increasing tortuosity for thicker interlayer}
    \label{fig:plots}  
\end{figure}

The gas-phase MacEtch targeting HAR structures must be carefully optimized. Here, we describe an example of etching condition optimization using a pattern with Hbar structures (200 nm lines) realized by the I-Et method with a 50 nm Cr interlayer. The procedure is similar for the other pattern transfer methods and structures with feature sizes in the range of 15-400 nm. It is worth noting that the goal of this optimization is to select the best etching conditions that can preserve the catalyst quality for a long etching process. MacEtch in gas phase has few parameters (HF solution composition, temperature, sample position, area of the sample) already analyzed in previous work, \cite{SHI2023107311} which can interplay, making the optimization very challenging. In the absence of a comprehensive model to guide the experimental procedure, we suggest here a minimal control matrix that can open the route to a more extensive and systematic work to assess the reproducibility of the process. \textbf{Figure \ref{fig:plots}}.a reports the etching rate as a function of temperature and HF refill measured in the center (average depth in a 10 $\upmu$m-wide region) of each sample after 15 min of MacEtch above HF[25\%]. A representative example of the SEM cross-section is shown in Figure \ref{fig:plots}.d. All the chips have the same size and originate from the same wafer. The chips were etched sequentially with short breaks between them. The difference between the samples was the temperature at which they were held during etching, which was performed in a randomly chosen order of 45, 55, 60, and 50 °C. The etching rate decreased with increasing etching temperature (Figure \ref{fig:plots}.a), which can be attributed to the reduced water adsorption on the surface. Water adsorption has been reported to catalytically support the HF removal of SiO$_2$ in the vapor-HF process.\cite{RN267} We can then assume that water reduction can also slow down the MacEtch process, despite MacEtch being thermally activated.\cite{RN278} This behavior was also observed in the shape of a volcano plot in a previous work.\cite{RN295} Furthermore, the comparison of etch rates for MacEtch in the gas phase is a very challenging task due to the difficult control of the reactant concentration. For small areas and short etches, the etchant consumption during etching can be neglected. Instead, the main loss occurs when the system is opened, and the HF vapor escapes the chamber. This was investigated by plotting the etching rate measured for identical samples under the same etching conditions at a temperature of 50 °C for 15 min above the HF [25\%] as a function of how often the system was opened after at least 15 min of evaporation (Figure \ref{fig:plots}.a). The high sensitivity of the etch rates to the HF concentration shows that controlling the etching conditions is crucial for reproducible results. This indicates the need to frequently replace HF to obtain comparable data from this setup.

The requirement for successful pattern transfer is a straight sinking of the catalyst without isolated deflection from the vertical direction into the substrate. This translates into a catalyst layer that remains intact at the end of the process. We analyzed the catalyst profile at the bottom of the etched trenches in the cross-sectional SEM images via the arc-length tortuosity and the etched angle with respect to the substrate (see Materials and Methods and supporting information for details). Tortuosity ($\tau$) quantifies the extent to which the catalyst deviates from a perfect profile ($\tau$=1). We propose the following quantitative metrics: a tortuosity of up to 1.1 represents good quality, and bad quality is assigned to the presence of multiple breaks or a large standard deviation in the etching angle (over 3°). Figure \ref{fig:plots}.d-f) show examples with tortuosities of 1.03, 1.05, and 1.23, respectively. Figure \ref{fig:plots}.b shows the etching angle and tortuosity as a function of temperature. There is no sensitive variation in the quality factors in this temperature range. However, we can observe a trend with increasing tortuosity and decreasing etching angle below 89° as the temperature increases to 60 °C. Considering the decrease in the etching rate as the temperature increased, we selected an etching temperature for the HAR structures in the range of 50-55 °C. 

In another example, we optimized the interlayer thickness. Figure \ref{fig:plots}.c reports an example of tortuosity measured for two different batches of chips with different interlayer thicknesses (Al$_2$O$_3$/SiO$_2$ with 30 nm/20 nm and 60 nm/20 nm, respectively) etched for 30 min with HF[50\%] at 55 °C. The batch with 80 nm-thick interlayer has a larger spread and more samples with $\tau$>1.1, indicating that chips with thinner interlayers are more reliable in terms of quality.

The etching conditions need to be optimized to achieve the best quality for each specific samples set by studying the SEM images in cross section at moderate depth (5-10 $\upmu$m). The etching rate of gas-MacEtch in this system is in the range of 0.1-0.9 $\upmu$m min$^{-1}$ for the used conditions of HF concentration and distance of the sample from the liquid. These etching rates are certainly lower than those of the conventional dry-etch with the Bosch process (3-8 $\upmu$m min$^{-1}$),\cite{RN621} but are in the same range as those of cryo-etching (0.6 $\upmu$m min$^{-1}$) for similar feature sizes.\cite{RN445} The main limitation of this system is the gas flow, especially because we cannot vary the O$_2$ flow, which would help to increase the etching rate. 

\begin{figure}
    \centering
    \includegraphics[width=1\textwidth]{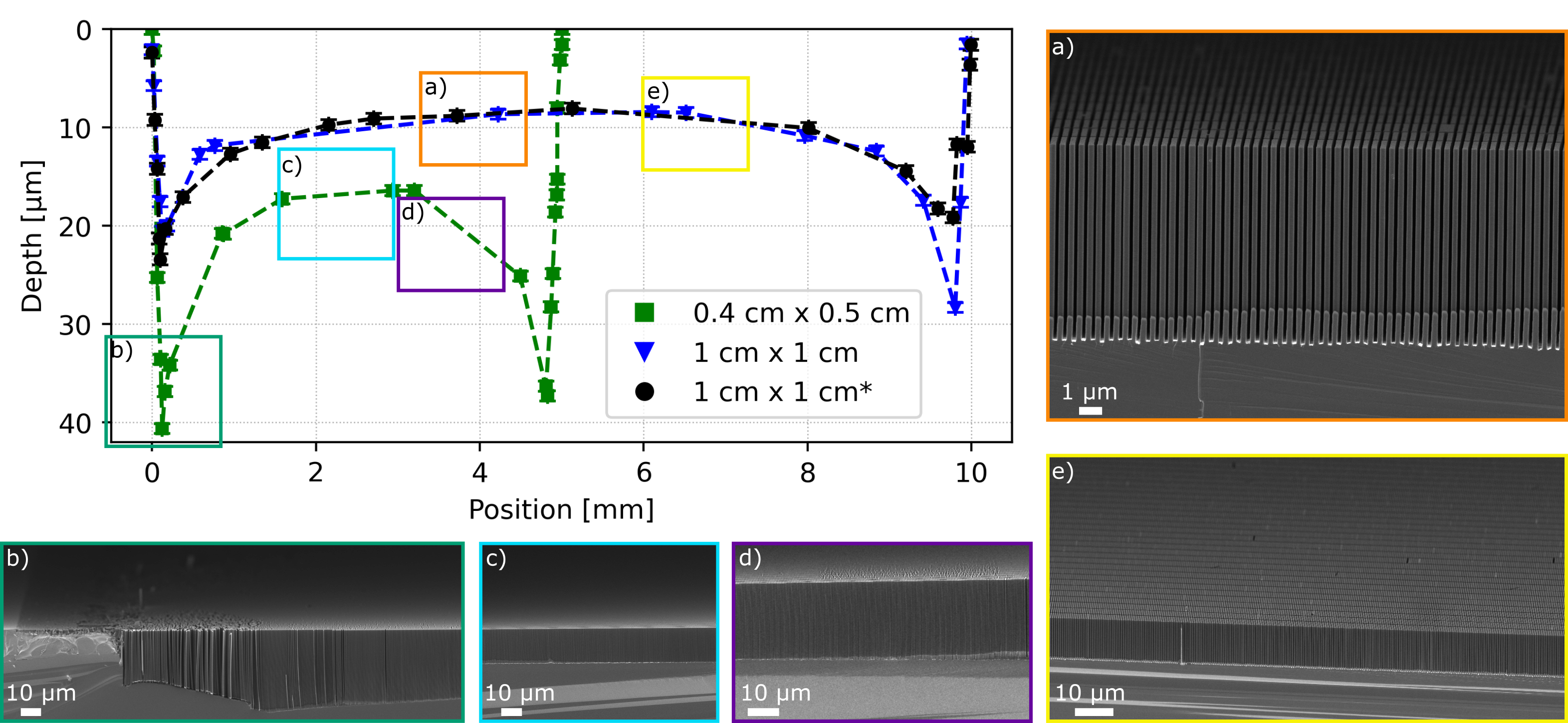}
    \caption{Etching depth of three samples of 200 nm line grating created with I-Et method (30 nm Cr) and subsequent MacEtch as a function of the position in the pattern. The dashed lines serve as guides for the eyes. The legend denotes the dimensions of the patterned area. All samples were etched under the same conditions: etching for 1 h at 50 °C over HF [25\%]. The sample marked with * was etched after 2 months. The depth profile was reconstructed by analyzing multiple SEM images at various positions. The squares show the selected SEM images in the cross-section at different locations.  }
    \label{fig:area}  
\end{figure}

\textbf{Figure \ref{fig:area}} shows the depth profile of a 200 nm line grating created with I-Et (30 nm Cr interlayer) and SEM images in cross-section at selected positions. The depth profile was obtained by carefully measuring the cross-sectional SEM images at a defined distance from the edge of the sample. The etched samples were patterned together on the same wafer, and the Cr interlayer was removed in a liquid etchant to enable Pt dewetting (at 350 °C for 1h, see previous work \cite{RN295}). In the first run, a 0.4 cm × 0.5 cm area sample was etched together with a 1 cm × 1 cm area sample. The second sample with a 1 cm × 1 cm area was etched two months later (see Supporting Information). All samples were etched for one hour at 50 °C over HF [25\%]. To make the results comparable, we used a freshly mixed solution for each run.

Etching is faster at the edges of the patterned areas, a sort of microloading effect due to a local depletion of the reactants in the region with high pattern density, also occurring in plasma etching.\cite{RN2988} The depth of the structures changed continuously throughout, with no relevant breaks in the catalyst apart from the edge. As the samples are patterned by I-Et, the outsides of the patterned area do not contain an active catalyst site. Hence, the unpatterned area does not consume reactants during MacEtch, resulting in a gradient of reactant concentration decreasing from the edges to the center of the pattern. Being an edge effect, it is sensitive to the sample area; in fact, the 1 cm × 1 cm sample (Figure \ref{fig:area}) has a more uniform etching in the center as the distance from the edge increases. The larger the pattern area, the larger the consumption of reactants, leading to an overall slower etch rate (loading effect), such that the larger sample reaches a 20–30 $\upmu$m etching depth at the edges with respect to the 40 $\upmu$m of the small sample. Similar edge effects have been reported for MacEtch in liquids\cite{RN381} and in gases.\cite{RN25, RN295} However, in L.L. Janavicius et al.\cite{RN25} both etchant and oxidant gases were flowing across the sample, with one side of the sample facing the flow (leading side). In our system, the HF evaporates perpendicular to the sample, whereas O$_2$ flows into the chamber from all the lateral sides of the sample (see Supporting Information). The etching profile appears symmetric, some minor differences at the edges can be related to the omission of the proximity effect correction in the lithographic step, which affects the first tens of $\upmu$m.\cite{czaplewski_holt_ocola_2013}. The decrease in the etch rate with the sample area could be counterbalanced by an increase in the reactant flow. Owing to the limitations of the tool, this is not further investigated here.
The overall etch profile was successfully replicated in a second experiment two months later, demonstrating the consistency of the experimental parameters, such as the replacement of the HF solution prior to each experiment and a clean pattern transfer. However, the later sample presents more defects (see supporting information). A third sample etched after storage for 8 months showed even more defects, supporting the claim of an aging effect, probably because Pt is subjected to cumulative carbon contamination.\cite{RN3154} These observations prompt the need for a study of catalyst degradation as a consequence of environmental conditions that go beyond the scope of this manuscript. 

\begin{figure}
    \centering
    \includegraphics[width=0.45\textwidth]{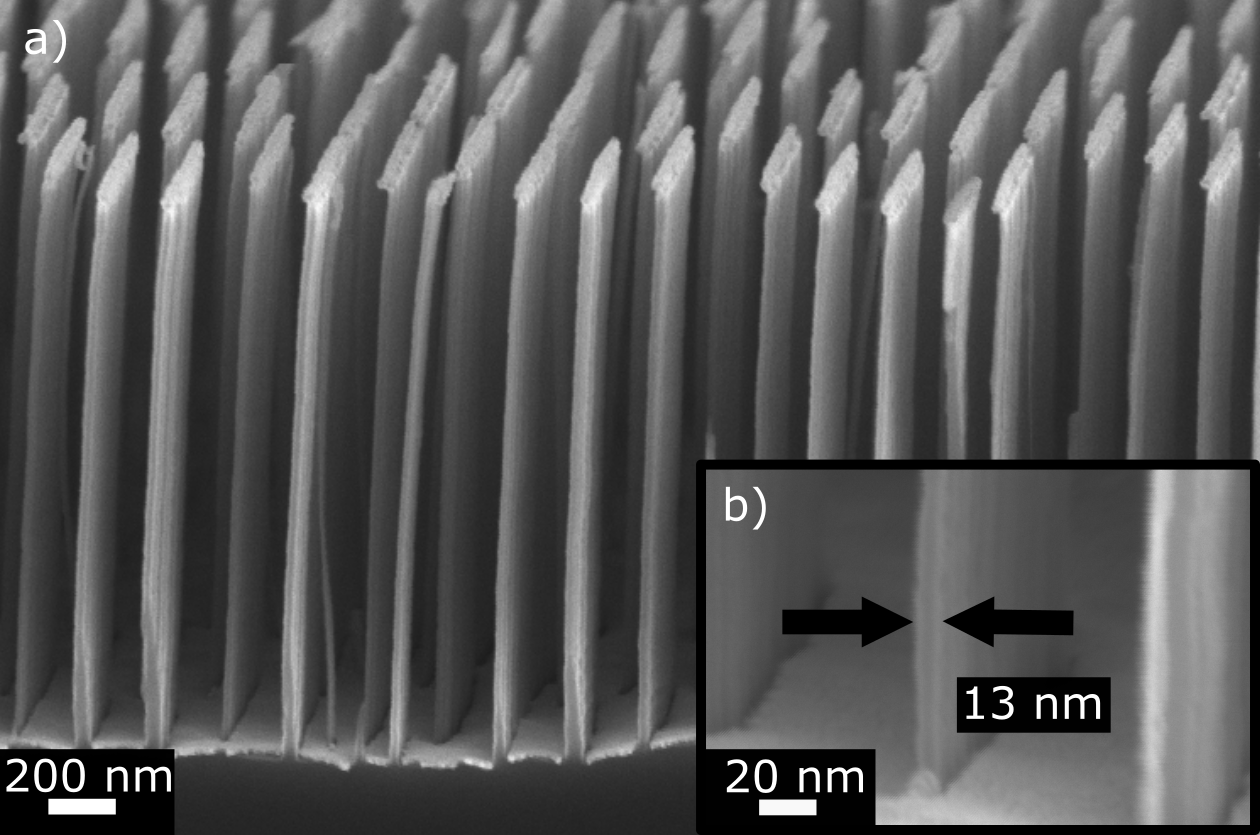}
    \caption{ a) Cross-sectional images of fin structures fabricated with I-Lo using Cr/Al$_2$O$_3$ and etched 15 min at 50 °C above HF [37.5\%]. b) Zoomed image of the structures, demonstrating fin width of 13 nm width and very little lateral etching.}
    \label{fig:Thin}  
\end{figure}

Using the unpatterned mesoporous catalyst, we estimated a feature size limit for gas-MacEtch of approximately 10 nm. In our previous work \cite{RN2627}, we observed very thin lamellae; however, lateral etching was predominant, compromising the structural stability of the fins. Herein, we report patterned structures with linewidths in the range of 10 nm realized by I-Lo and an advanced undercut of the Cr/Al$_2$O$_3$ (30nm/20nm) interlayer. The structure is very sensitive to the etching conditions (see \textbf{Figure \ref{fig:Thin}}.a) because even slight sideways etching was sufficient to release the structure. The lamellae are as thin as 13 nm (see Figure \ref{fig:Thin}.b). The Pt catalyst at the bottom of the fins shows a gap to the sidewall in the order of nanometers, indicating almost negligible lateral etching of MacEtch in the gas phase. This hints that feature size in the range of 10 nm or lower are feasible, in line with previous observations in wet-MacEtch.\cite{RN223}

\subsection{High aspect ratio structures }
\begin{figure}
    \centering
    \includegraphics[width=1\textwidth]{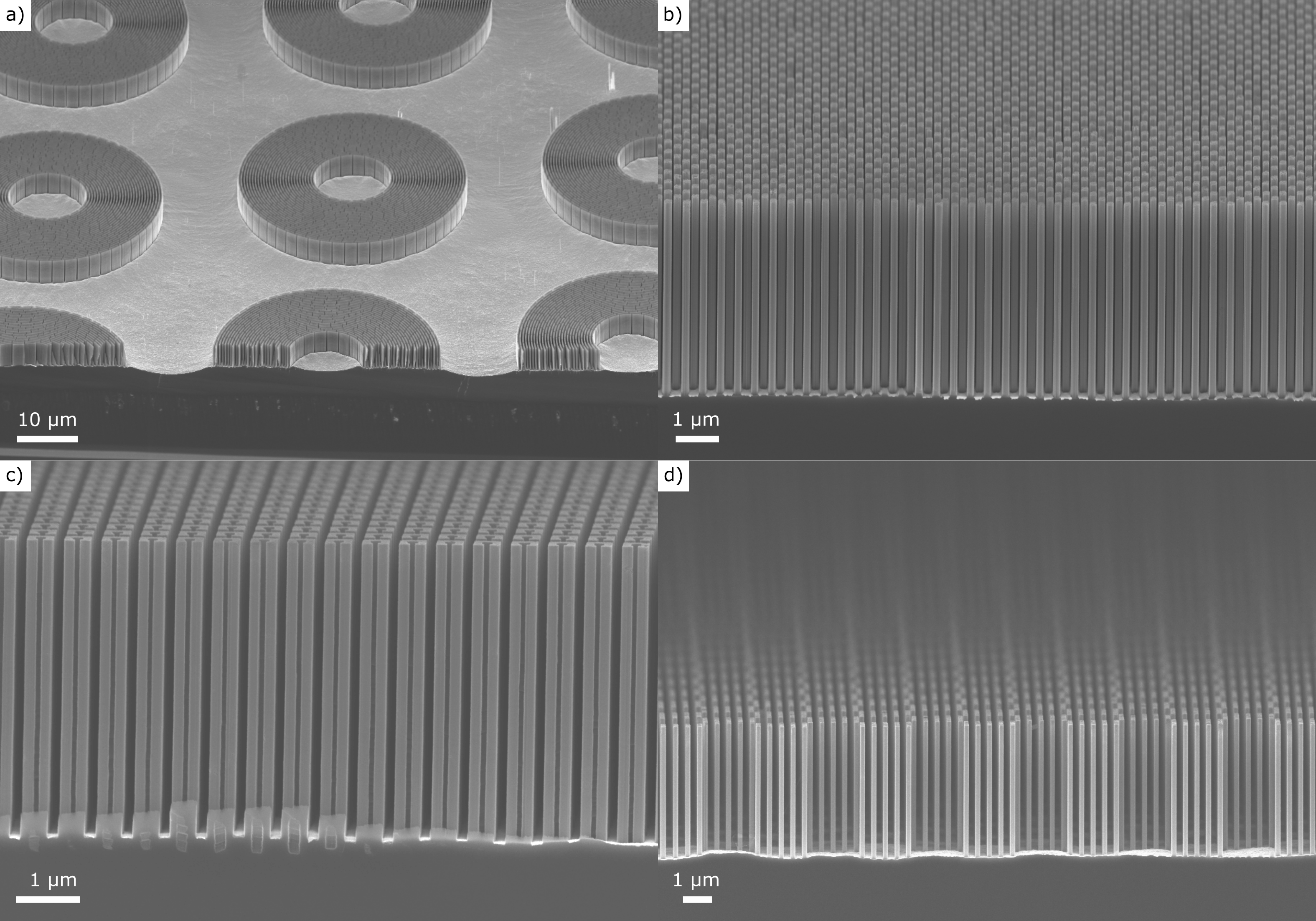}
    \caption{ Cross-sectional SEM of various structures created with the I-Lo method (30 nm Al$_2$O$_3$/20 nm SiO$_2$) etched for 30 min above HF [37.5\%]. a) ring grating etched at 50 °C  with a small pitch of 460 nm and large pitch of 50.6 $\upmu$m (SEM tilt of 30°); b) checkerboard grating etched at 50 °C (SEM tilt of 30°); c) ring-oscillator structures etched at 45 °C with an opening of 85 nm and pitch of 300 nm; d) double checkerboard etched at 50 °C with double pitch(P1=400 nm and P2=3.6 $\upmu$m). }
    \label{fig:different}  
\end{figure}

Metal-assisted chemical etching can etch samples to considerable depths, regardless of feature size. The improved patterning process with more consistent etching allows for a large etching depth in high aspect ratio structures with a high level of control. Both the I-Et and I-Lo methods can be used to create high aspect ratio structures. 
In \textbf{Figure \ref{fig:different}} we present a selection of optical elements that highlight the capability and flexibility of MacEtch. The conventional fabrication of these structures at this aspect ratio is challenging or impossible with more established etching methods, highlighting the advantages of MacEtch. A tapering profile with vertical angle far below 89° will be produced by dry-etch with Bosch process in dense submicron gratings, while the large open areas will be much more deep using plasma process due to the microloading effect.\cite{RN2982} 
These structures can all be etched with a higher aspect ratio and functional tops, given sufficient optimization. Diffractive Optics in the X-ray regime usually require highly precise ordered vertical structures with sub-micron feature sizes and high aspect ratios,\cite{Rawlik:23,RN376} owing to the high transparency of materials to X-rays. Circular gratings are regularly used in X-ray optics in the form of Fresnel zone plates \cite{vila-comamala_jefimovs_raabe_kaulich_david_2008} with varying zone widths or in arrays, as here, to produce a phase variation in the wavefront, which is useful to detect directional features, such as in X-ray tensor tomography. \cite{RN53} Checkerboard patterns with single \cite{Morgan:13} (Figure \ref{fig:different}.b) or multiple pitches \cite{RN2989} (Figure \ref{fig:different}.d) are used as optical elements in grating interferometry. The presented structures are all created with I-Lo (30 nm Al$_2$O$_3$/20 nm SiO$_2$) and etched above HF[37.5\%] for 30 min at various temperatures. Figure \ref{fig:different}.a shows an array of circular gratings with a small pitch of 460 nm and a large pitch of 50.6 $\upmu$m. Figure \ref{fig:different}.b shows a checkerboard grating with a pitch of 400 nm. Figure \ref{fig:different}.c shows a ring resonator with a pitch of 600 nm and an opening of 85 nm, demonstrating a structure with a hollow center.  Figure \ref{fig:different}.d shows a double checkerboard grating with a small pitch P2 of 400 nm and a large pitch P1 of 3.6 $\upmu$m.

\begin{figure}
    \centering
    \includegraphics[width=1\textwidth]{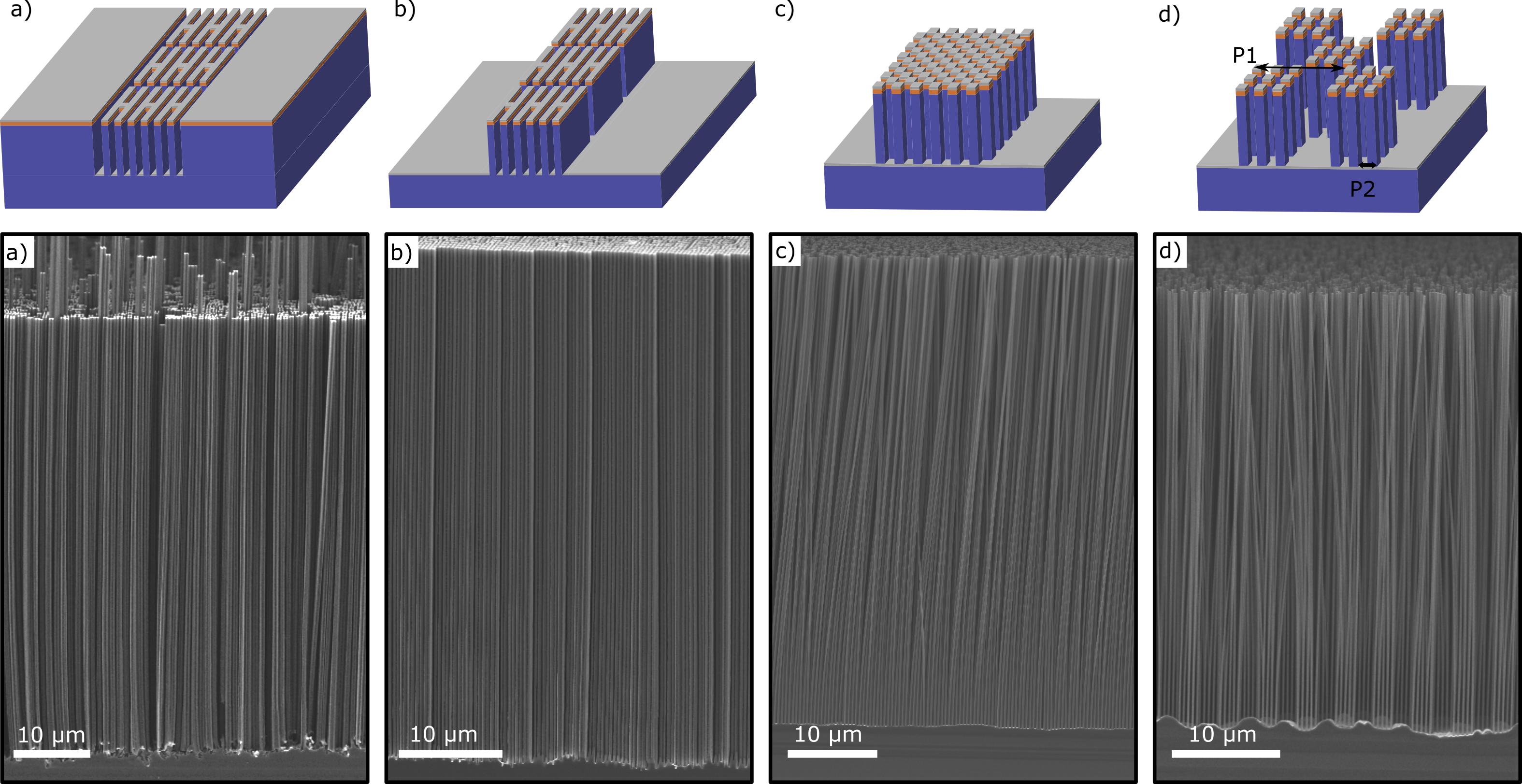}
    \caption{ Different schematics and cross-sectional SEM images of high aspect ratio structures. a) H-bar grating created with I-Et (50 nm Cr), the sample was etched 3h at 55 °C above HF [12.5\%]; b) H-bar grating created with I-Lo (50 nm Cr), the sample was etched 30 min at 50 °C above HF [50\%]; c) checkerboard grating created with I-Lo (30 nm Al$_2$O$_3$/20 nm SiO$_2$), the sample was etched 60 min at 50 °C above HF [50\%]; d) double checkerboard grating created with I-Lo (30 nm Al$_2$O$_3$/20 nm SiO$_2$), the sample was etched 60 min at 50 °C above HF [50\%].}
    \label{fig:aspect_ratio}  
\end{figure}
\textbf{Figure \ref{fig:aspect_ratio} }shows cross-sectional SEM images of various structures with a feature size of approximately 200 nm and a depth of approximately 50 $\upmu$m created with I-Et and I-Lo. Shown are a H-bar structure (linewidth 200 nm) created by I-Et that was etched for 3h at 55 °C above HF [12.5\%] (Figure \ref{fig:aspect_ratio}.a); an H-bar structure (linewidth 200 nm) created by I-Lo, etched for 30 min at 50 °C above HF [50\%] (Figure\ref{fig:aspect_ratio}.b); a checkerboard pattern (linewidth 200 nm), created by I-Lo, etched for 60 min at 50 °C above HF [50\%] (Figure \ref{fig:aspect_ratio}.c); an I-Lo double checkerboard grating (P1=400 nm and P2=3.6$\upmu$m) etched for 60 min at 50 °C above HF [50\%].
Both the I-Et and I-Lo patterning methods can etch 200 nm linewidth h-bar structures to a depth of 50 $\upmu$m. The checkerboard pattern shows the possibility of deep etching (53 $\upmu$m) with a 50\% dense structure. The double checkerboard can etch deep structures (40 $\upmu$m) with masks of two different sizes in the same pattern. The wavy catalyst pattern at the bottom of the trenches indicates a very minor dependence of the etch rate on the pitch size (Figure \ref{fig:aspect_ratio}.d).
The achieved etching depth and the resulting aspect ratio of over 250:1 are in the same regime as the self-assembled pattern from Pt dewetting.\cite{RN295} The structure all etch predominantly straight with smooth sidewalls. This shows a clear advantage over etching in plasma, in which sidewall roughness, tapering, and aspect-ratio-dependent etching rates, such as ARDE and RIE-lag, are well-known drawbacks.\cite{RN2982} In particular, for applications such as meta-lenses\cite{baracu2021,RN77} and photonic nanocavities \cite{RN2985}, where a uniform structure height and sidewall smoothness are needed for a variety of different feature sizes, MacEtch in the gas phase can be an attractive alternative. 

Some of the H-bar structures in the sample processed using the I-Et method appear extruded (Figure \ref{fig:aspect_ratio}.a). In this example, the catalyst exhibited uneven and uncontrolled etching at the bottom of the grating trench. The etching conditions of this sample were set at a low HF concentration (HF [12.5\%]) to reduce the etching rate, as faster etching is not sustainable for this I-Et pattern. MacEtch has been reported to etch different crystallographic directions, and the switching from the main etching direction of $\langle 100 \rangle$, which is the wafer surface, is mainly dictated by the etching rate\cite{RN278}. If the etch front reaches this regime, for example, by locally unbalancing the HF/O$_2$  ratio, etching proceeds in a different direction and chaotic etching occurs, which breaks the catalyst. This occurred at the bottom of the structure after a straight etching of approximately 50 $\upmu$m, and the structure was subsequently extruded. Assuming a high HF concentration, the catalytic decomposition of the oxidant is the rate-determining step. In the I-Et (Figure \ref{fig:aspect_ratio}.a) sample, the area external to the pattern blocked the catalyst, and the active catalyst area was 50 times smaller than that of the I-Lo sample. Similar to the observations regarding the loading effect (see Figure \ref{fig:area}), the difference in the active area of the catalyst causes a relevant difference in the consumption of the reactants, and hence a different etch speed. Despite the lower HF concentration in the I-Et sample and the etching rate being three times slower with respect to the I-Lo sample, the diffusion through the 50 $\upmu$m tall structures altered the HF/O$_2$ balance at the bottom, which triggered the onset of etching events in different directions. The supply of reactants becomes a limiting factor in ultra-high aspect ratios due to diffusion, indicating the need for a system with gas flow control. \cite{RN381} In comparison, the samples processed by I-Lo were etched with a much higher nominal HF concentration and exhibited an intact catalyst with no structural damage, indicating that the dynamic consumption of reactants operated by the larger area of exposed catalyst (Pt is active also outside of the patterned area) is more efficient in optimizing the etchant concentration. These examples show the need to adjust the etching concentration for the successful etching of each specific pattern and catalyst active area.

\section{Conclusion}
In this study, we present a novel patterning method using EBL to improve catalyst preparation for MacEtch in the gas phase of high aspect ratio nanostructures. A crucial innovation is the introduction of an interlayer material that physically separates the catalyst from the lithographically processed resist, reducing catalyst contamination and allowing for a more thorough cleaning before Pt deposition. We demonstrated two different approaches for pattern transfer in the catalyst layer in the presence of an interlayer material, using plasma-based pattern transfer (I-Et) or lift-off (I-Lo). The examples cover a wide range of structures with feature sizes of 10-400 nm and different materials with interlayer capability (Cr, Al$_2$O$_3$ and SiO$_2$). The proposed method paves the way for the creation of elongated Si nanostructures with a high aspect ratio while retaining a multilayer top coating via MacEtch. 
The criteria for etching optimization focused on the structural quality of the catalyst at the bottom of the trenches, evaluated by analyzing the etching angle and the tortuosity in SEM images in cross section. The depth profile was carefully measured by sampling the etched depth at different positions, revealing a microloading effect with edges etching faster owing to a local depletion of the reactants,  which is the main limiting factor of the HF evaporation tool at atmospheric pressure. Real-time detection of actual reactant consumption and the possibility of regulating gas flow in a more sophisticated tool could open new possibilities and optimization strategies. Nanofabrication with aspect ratios exceeding 250:1 has been demonstrated in a variety of optical elements useful for applications in the X-ray regime using both the I-Lo and I-Et approaches. The extreme sidewall smoothness and verticality of the etched features are relevant for controlling optical manipulation in nanophotonics. Our findings help understand the relevance of interface chemistry and the fundamental role of catalyst preparation for MacEtch of silicon, contributing to the development of the necessary building blocks to exploit the plasma-free nanostructuring of MacEtch in the gas phase.

\section{Materials and Methods}

Because the goal is to demonstrate the fidelity of pattern transfer during MacEtch, especially in the direction perpendicular to the substrate, we designed the patterning unit with a functional shape: a large mesh of an interconnected catalyst\cite{RN295} with a periodic array \cite{RN201} of units so that the eventual deflection due to the catalyst motion can be easily visualized at low magnification as a defect in the periodicity; a pattern unit that produces silicon pillars with a cross-section of an H bar or rectangles, so that the corners can be easily detectable in cross-section scanning electron microscopy (SEM) with minimal deviation in footprint with respect to circular pillars, which are the most used MacEtch structures. H bars are also more stable in terms of stiction with respect to rectangles in the case of liquid condensation during long etching processes or humidity before the SEM inspection; therefore, this type of functional pattern is more suitable for investigating very high aspect ratio features. 

All samples were prepared on fresh N-type phosphor-doped 4-inch silicon wafers with a $\langle 100 \rangle$ orientation, resistivity of 1-30 M$\Omega$ cm, and wafer thickness of 250 µm. Prior to patterning, the wafers were cleaned using oxygen plasma. Cr, Al$_2$O$_3$, SiO$_2$ and Pt were deposited using electron beam evaporation with a BakUni system from Evatec. The Pt thickness was 12 nm in all samples, and the thickness of the interlayer was in the range of 20 to 80 nm (See Table 1). Lithography was performed by electron beam lithography using a Raith EBPG 5000Plus and the positive-tone resist CSAR-62 from Allresist GmbH.  The dimensions of the patterned area were 100 $\upmu$m × 2 mm, unless mentioned otherwise. This small size reduces the writing time while allowing for consistent dicing of the structure along its long side. Larger patterns were produced to test uniformity (Figure \ref{fig:area}). The developer AR 600-546 and remover AR 600-71 from Allresist GmbH were used for the resist owing to their high resolution and fast removal. For the I-Lo process, we used a process adapted from Ohlin et al.\cite{RN363} which consists of an initial soak followed by sonication. The substrates were cleaned using coupled plasma reactive ion etching with O$_2$. This removes the remaining resist and etches Si, further aiding in the separation of the layers. Cr was etched with Cl$_2$/O$_2$ plasma or diluted liquid Technietch Cr 01 from Microchemicals GmbH (the samples were immersed in the solution and agitated in an ultrasonic bath.); Al$_2$O$_3$ was etched selectively over SiO$_2$ in a 3:1 liquid solution of H$_2$SO$_4$ and H$_2$O$_2$. 
Thermal dewetting of platinum was performed on a metal-covered hotplate in air for 1h. 
The gas-phase MacEtch process consists of exposing the Pt-patterned Si samples to HF vapor and O$_{2}$ gas, as described in a previous work.\cite{RN295} During etching, the sample is held in place and in contact with a heated holder at a fixed temperature by physical clamping. For etching, the basin was filled with aqueous HF. This created an etching chamber filled with air and HF vapor at ambient pressure.  MacEtch was performed using the vapor HF tool developed by Idonus (see supporting information). The liquid solution had a fixed total volume of 200 ml and the composition was varied by mixing different volumes of DI water (18 M$\Omega$) and aqueous HF from Technic at 50\% volume concentration. 
The samples were mechanically clamped on the heating holder and held 1.3 cm above the liquid to be exposed to HF vapor. The etching time was determined as soon as the liquid began to flow. One minute before stopping the timer, the liquid flowed out of the chamber, and the process was stopped by removing the top of the sample. The sample was then kept at this temperature to allow the HF to evaporate from the sample for safety. 
The main parameters for this experiment consisted of the holder temperature, giving the sample temperature; the HF concentration in the liquid, giving the HF concentration in the gas; the distance between the sample and the liquid, affecting the vapor concentration at the sample surface; and the etching time. The first three parameters affect the chemistry of the process and hence the reaction rate, as shown in previous studies.\cite{SHI2023107311} The etch rate increases as a function of HF concentration, presenting a volcano plot as a function of temperature, with a maximum at approximately 45 °C, and decreases as a function of the distance of the sample from the liquid level. In this study, the etching conditions were optimized by varying the temperature in the range of 45-60 °C and the amount of HF 50\% in the liquid solution. The total amount of liquid was fixed at 200 ml, which was a mixture of HF and DI. We use a compact notation for the HF concentration; for example, HF [50\% ] corresponds to 200 ml HF 50\%, HF [25\%] corresponds to 100 ml HF 50\% and 100 ml DI, and so on. The etching optimization criteria focused on the structural quality.

Cross-sectional SEM inspection was performed by cleaving the silicon samples.  SEM imaging was performed using a Zeiss Supra VP55 microscope under a 10° tilt, unless otherwise mentioned, with active tilt correction.

The quantitative analysis used to guide process optimization consists of three separate quantities. The etch rate is calculated from the measured etch depth and etch duration. The measurements were carried out on one sample each, where the given depth was probed at multiple locations in the center of the region, and the average and standard deviation of this measurement were taken together with a conservative estimate of SEM uncertainty of 100 nm. The angle measurement is the average etch angle. This was obtained by measuring the angle of the top surface of the structure and the right side, where the line from the corner of the top structure to the etch front was taken. The given value is the average and standard deviation of multiple measurements within the specified range. The surface tortuosity was quantified as the ratio of the manually traced profile length of the catalyst to the straight-line distance across the cross-section. The main error is acknowledged to be from the manual tracing and was estimated by multiple tracings as being at 1$\%$ of the reported value. To quantify the deformation of the etch front, the Pt profile was approximated by taking a point every 100 nm (see supporting information).

The MacEtch conditions used in this study are listed in \textbf{Table \ref{Tab:cond}}. We remind the reader that these data are only suited for broad comparison, as the concentration of the liquid and the overall patterned area vary between samples.

\begin{table}
\centering
\caption{MacEtch conditions and characteristics of the samples presented in this work and the relative Figures: thickness (\textit{d}) and materials of the interlayer for patterning in the order of deposition; HF 50\% volume in ml; temperature (\textit{T}) of the sample during etching; etching time (\textit{t}); size (\textit{S}) of the features in the pattern; measurement of the etching depth in the middle of the patterned area; angle of the etched structures with respect to the substrate; and Tortuosity ($\tau$) with a standard error of $1\%$ measured at the Pt etch front.}
\begin{tabular}{||l c l r c c r r r l||} 
 \hline
Figure & Method & \textit{d} [nm] & HF  & \textit{T} [°C] & \textit{t} [min] & \textit{S} [nm] & Depth [um] & Angle [°] & $\tau$ \\
 \hline\hline
 \ref{fig:interlayer_etching}.f & I-Et & 50 Cr & 100  & 55 & 15 & 200 & 4.91 $\pm$ 0.04 & 90.2 $\pm$ 0.92 & 1.01\\ 
 \hline
 \ref{fig:interlayer_etching}.g,h & I-Et & 50 Cr & 100  & 50 & 15 & 400 & 7.50 $\pm$ 0.18 & 89.0 $\pm$ 3.0\phantom{0} & 1.1\\ 
 \hline
 \ref{fig:interlayer_LO}.f & I-Lo & 50 Al$_2$O$_3$& 200 & 60 & 30 & 400 & 14.93 $\pm$ 0.18 & 89.48 $\pm$ 0.13 & 1.01 \\ 
 \hline
 \ref{fig:profile}.b & I-Lo & 30 Al$_2$O$_3$& 100  & 45 & 30 & 400 & 0.24 $\pm$ 0.17 & 90.5 $\pm$ 2.5\phantom{0} & 1.01  \\
 \hline
 \ref{fig:profile}.e & I-Lo & 30 Al$_2$O$_3$& 200  & 55 & 10 & 230 & 3.48 $\pm$ 0.13 & 92.4 $\pm$ 1.4\phantom{0} & 1.04 \\
 \hline
 \ref{fig:bilayer}.e & I-Lo & 30 Cr + 20 Al$_2$O$_3$& 150  & 55 & 15 & 200 & 1.32 $\pm$ 0.05 & 87.5 $\pm$ 1.2\phantom{0} & 1.03 \\
 \hline
 \ref{fig:plots}.d & I-Et & 50 Cr& 100  & 55 & 15 & 200 & 4.81 $\pm$ 0.06 & 90.41 $\pm$ 0.94 & 1.02 \\
 \hline
 \ref{fig:plots}.e & I-Et & 50 Cr& 100  & 55 & 15 & 400 & 3.88 $\pm$ 0.06 & 90.9 $\pm$ 2.0\phantom{0} & 1.05  \\
 \hline
 \ref{fig:plots}.f & I-Et & 50 Cr& 100  & 50 & 15 & 400 & 3.88 $\pm$ 0.06 & 92.4 $\pm$ 9.8\phantom{0} & 1.23  \\
 \hline
 \ref{fig:area}.a,e & I-Et & 30 Cr& 100  & 50 & 60 & 200 & 8.52 $\pm$ 0.05 & 89.26 $\pm$ 0.14 & 1.03  \\
 \hline
 \ref{fig:area}.b,c,d & I-Et & 30 Cr& 100  & 50 & 60 & 200 & 16.59 $\pm$ 0.09 & 90.31 $\pm$ 0.15 & 1.03 \\
 \hline
  \ref{fig:area}.* & I-Et & 30 Cr& 100 & 50 & 60 & 200 & 8.06 $\pm$ 0.11 & 90.0 $\pm$ 1.4\phantom{0} & 1.05\\
 \hline
 \ref{fig:Thin} & I-Lo & 30 Cr + 20 Al$_2$O$_3$& 150  & 50 & 15 & 15 & 1.61 $\pm$ 0.04 & 88.4 $\pm$ 1.2\phantom{0} & 1.02  \\
 \hline
 \ref{fig:different}.a & I-Lo & 30 Al$_2$O$_3$+ 20 SiO$_2$& 150  & 50 & 15 & 230 & 3.45 $\pm$ 0.11 & 89.6 $\pm$ 1.8\phantom{0} & 1.03 \\
 \hline
 \ref{fig:different}.b & I-Lo & 30 Al$_2$O$_3$+ 20 SiO$_2$ & 150  & 50 & 15 & 200 & 4.46 $\pm$ 0.04 & 89.64 $\pm$ 0.13 & 1.01  \\
 \hline
 \ref{fig:different}.c & I-Lo & 30 Al$_2$O$_3$+ 20 SiO$_2$ & 150  & 45 & 15 & 150 & 4.73 $\pm$ 0.04 & 89.09 $\pm$ 0.16 & 1.01  \\
 \hline
 \ref{fig:different}.d & I-Lo &30 Al$_2$O$_3$+ 20 SiO$_2$ & 150  & 50 & 15 & 200 & 4.59 $\pm$ 0.05 & 89.93 $\pm$ 0.14 & 1.01 \\
 \hline
 \ref{fig:aspect_ratio}.a & I-Et & 50 Cr& 50  & 55 & 180 & 200 & 53.46 $\pm$ 0.50 & 89.02 $\pm$ 0.93 & 1.37  \\
 \hline
 \ref{fig:aspect_ratio}.b & I-Lo & 50 Cr& 200  & 50 & 30 & 200 & 49.48 $\pm$ 0.25 & 89.58 $\pm$ 0.45 & 1.07  \\
 \hline
 \ref{fig:aspect_ratio}.c & I-Lo & 30 Al$_2$O$_3$+ 20 SiO$_2$& 200  & 50 & 60 & 200 & 52.33 $\pm$ 0.45 & - & 1.01 \\
 \hline
 \ref{fig:aspect_ratio}.d & I-Lo & 30 Al$_2$O$_3$+ 20 SiO$_2$& 200  & 50 & 60 & 200 & 39.77 $\pm$ 0.24 & - & 1.10\\ [0ex] 
 \hline
\end{tabular}
\label{Tab:cond}  
\end{table}

\medskip
\textbf{Supporting Information} \par
Supporting Information is available from the Wiley Online Library or from the author.

\newpage
\medskip
\textbf{Acknowledgements} \par
The authors thank the funding agencies: Swiss Nanoscience Institute; the SNI PhD grant MAGNET; PHRT-TT Project Nr. 2022-572 INTIMACY; SNSF grant Nr. 200021-232339 ABSOLUTE; PROMEDICA Stiftung Chur, Grant Nr.  1788/M. We thank: the PSI cleanroom facilities PICO; K. Jefimovs, G. Mihai, D. Marty, V. A. Guzenko, C. Wild and A. Weber (PSI), who provided technical support; A. H. Oechsle and Prof. T. Jung (PSI) for helping with XPS measurements; Prof. M. Poggio at the University of Basel for PhD supervision. 

\textbf{Conflicts of interest} \par
The authors have no conflicts of interest to declare.

\textbf{Data Availability Statement} \par
Data supporting the findings of this study are available from the corresponding author upon reasonable request.

\medskip

\bibliography{References}

\newpage
\begin{figure}
\textbf{Table of Contents}\\
\medskip
  \includegraphics{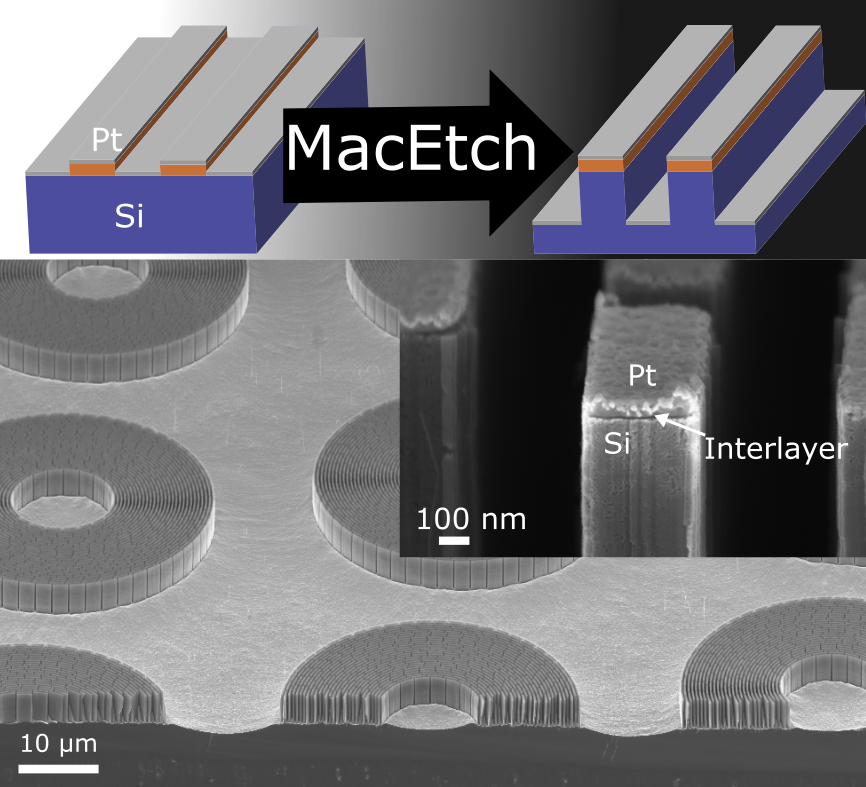}
  \medskip
  \caption*{An advanced patterning method is presented for metal assisted chemical etching. The interlayer replaces the resist and enables catalyst deposition on a clean interface. The prepared samples show improved performance with a more reproducible etching profile, higher aspect ratio, and increased consistency over a larger area. This study demonstrates the relevance of the catalyst-substrate interface. }
\end{figure}

\end{document}